\documentclass[conference]{IEEEtran}
\IEEEoverridecommandlockouts

\usepackage{cite}
\usepackage{amsmath,amssymb,amsfonts}
\usepackage{algorithmic}
\usepackage{graphicx}
\usepackage{textcomp}
\usepackage{xcolor}
\usepackage{subcaption}
\usepackage{supertabular,enumitem}
\usepackage{url}
\usepackage{diagbox}
\usepackage{booktabs} 
\usepackage{array} 

\usepackage{tcolorbox}
\usepackage{amsmath}
\usepackage{lipsum} 

\newcounter{o}

\usepackage{longtable} 
\def\BibTeX{{\rm B\kern-.05em{\sc i\kern-.025em b}\kern-.08em
    T\kern-.1667em\lower.7ex\hbox{E}\kern-.125emX}}

\usepackage{amssymb}
\newboolean{showcomments}
\setboolean{showcomments}{true} 
\ifthenelse{\boolean{showcomments}}
{\newcommand{\nb}[2]{
		\fcolorbox{gray}{yellow}{\bfseries\sffamily\scriptsize#1}
		{$\blacktriangleright$#2$\blacktriangleleft$}
	}
	
}
{\newcommand{\nb}[2]{}
	 
}

\IEEEoverridecommandlockouts \IEEEpubid{\makebox[\columnwidth]{ 979-8-3503-7128-4/24/\$31.00~\copyright2024 IEEE\hfill} \hspace{\columnsep}\makebox[\columnwidth]{ }}

\usepackage{hyperref}
    
\begin{document}

\title{

What Do Developers Discuss in Their Workplace? An Analysis of Workplace StackExchange Discussions\\
}

\author{\IEEEauthorblockN{ Natasha Grech} 
\IEEEauthorblockA{\textit{Trent University} \\
\textit{Peterborough, ON}\\
 Canada   \\
\textit{natashagrech@trentu.ca}}

\and

\IEEEauthorblockN{Md Farhad Hossain} 
\IEEEauthorblockA{\textit{Trent University} \\
\textit{Peterborough, ON}\\
 Canada   \\
\textit{mdfarhadhossain@trentu.ca}}

\and
\IEEEauthorblockN{ Omar Alam} 
\IEEEauthorblockA{\textit{Trent University} \\
\textit{Peterborough, ON}\\
 Canada   \\
\textit{omaralam@trentu.ca}}

}
\maketitle

\begin{abstract}

Software workplaces are increasingly recognized as key spaces for professional development, where developers encounter various challenges in their roles, which they often discuss in online forums. This paper analyzes 47,368 posts on the Workplace StackExchange site, aggregating developer insights and applying topic modeling techniques. Through manual analysis, we identified 46 distinct topics grouped into seven categories: Employee Wellness, Communication, Career Movement \& Hiring, Conflicts \& Mistakes, Corporate Policies, Management/Supervisor Responsibilities, and Learning \& Technical Skills. Our findings show that approximately 30\% of discussions involve workplace conflicts, marking this as the most prominent topic. Additionally, we found that workplace culture, harassment, and other corporate policy-related issues represent significant areas of difficulty commonly discussed among developers.

\end{abstract}


\section{Introduction}

The software industry has experienced unprecedented growth over the past few decades, driven by rapid technological advancements and an increasing reliance on digital solutions across all sectors. This expansion has resulted not only in a rise in the number of software companies but also in a transformation of software workplaces, compelling organizations to adapt to new workforce dynamics and operational models \cite{cusumano2021elements}. 

As software becomes central to nearly every industry, companies must support a workforce capable of keeping up with technological advancements. This has heightened challenges around skill development, employee well-being, and retention \cite{davis2019evolving}. Additionally, as software companies grow and globalize, creating inclusive and culturally adaptable workplaces has become a priority. Companies increasingly recognize that diversity drives innovation and enhances problem-solving, leading to initiatives such as mentorship programs for women, flexible work policies, and equal opportunity efforts \cite{smith2020fostering, chen2018diversity}.

Online software workplace discussions have become integral to how professionals in the software industry collaborate and share knowledge. As the software landscape evolves rapidly, these discussions facilitate continuous learning, problem-solving, and networking among developers, engineers, and other stakeholders. Moreover, online discussions play a crucial role in addressing workplace-related topics, such as team dynamics, conflict resolution, and career advancement strategies. Engaging in these discussions allows software professionals to better navigate workplace complexities and improve job satisfaction \cite{hill2021promoting}. As organizations increasingly adopt remote work and flexible arrangements, online forums play a crucial role in maintaining collaboration and cohesion across distributed teams.


Motivated by this, we conducted a large-scale analysis of developer online discussions on workplace-related issues. Specifically, we analyzed 47,368 posts on the Workplace StackExchange site, a question-and-answer platform dedicated to workplace and career topics. Workplace is a large online community where thousands of developers ask and answer questions on various issues, including job hunting, interviewing, salary negotiation, and professionalism in the workplace \cite{theWorkplace}. We aggregated developer discussions and applied topic modeling techniques. Workplace is part of the Stack Exchange network, a broad online community where millions of developers engage in Q\&A discussions.

In this study, we address the following four research questions:

\textbf{RQ$_1$ What workplace topics are discussed on the Workplace Stack Exchange?}
We find that developers discuss a wide range of workplace-related topics. Their discussions on Stack Exchange's Workplace encompass 46 different topics, including employment contracts, technical skills, team organization, work conflicts, workplace satisfaction, communication tips, and company policies. Our analysis reveals that work conflicts account for the highest number of posts, followed by resignation and job change. These topics belong to seven categories, Employee Wellness, Communication, Career Movement \& Hiring, Conflicts \& Mistakes, Corporate Policies, Management/Supervisor Responsibilities, and Learning \& Technical Skills. 
    
\textbf{RQ$_2$ How do the topics evolve over time?}
We find that topics related to conflicts and mistakes, as well as career movement and hiring, have the highest number of posts and consistently maintain the highest levels of activity over the years compared to other categories.

\textbf{RQ$_3$ How do the topics vary in terms popularity and difficulty? }
 We observe that topics vary significantly in popularity. \textit{Work Conflicts} is the most popular topic, with the highest number of views, while the technical topic \textit{Data Science} is the least popular and also the most difficult, as its questions take the longest to receive an accepted answer. \textit{Work Culture} and \textit{Harassment} are also among the most difficult topics, whereas \textit{New Graduate Jobs} and \textit{Languages} are the least difficult.

 \textbf{RQ$_4$ How does user activity and gender distribution vary across The Workplace discussion community? }

Since Workplace is a community of users interested in discussing workplace-related issues, we aim to understand user characteristics. We found that the top 20\% of users contribute to 17.2\% of the total posts in the dataset. Additionally, we found that men significantly outnumber women, with about 14\% of identified users being women and 86\% being men.

The remainder of this paper is organized as follows. The next section discusses related work. Section 3 covers the methodology and data collection. Section 4 presents the results related to the research questions. Section 5 discusses threats to validity. Section 6 concludes the paper.

\section{Related Work}

Stack Exchange \cite{stackExchange} is a key resource for topic modeling research, offering insights into software development practices through developer discussions. Many studies have utilized Stack Exchange. Tahir et al. \cite{tahir2020} utilize topic modeling on posts from Stack Overflow (SO) \cite{stackOverflow}, Software Engineering Stack Exchange \cite{softwareEng}, and Code Review Stack Exchange \cite{codeReview} to analyze developer concerns about code smells and anti-patterns. Chakraborty et al. \cite{chakraborty2021} investigate discussions on popular programming languages—Swift, Go, and Rust—showing how their emergence shapes developer discourse. Santos et al. \cite{santos2023} analyze technical debt discussions on the Stack Exchange Project Management \cite{projectManagement} site, using manual qualitative analysis rather than topic modeling. Other studies have applied topic modeling on SO for topics such as Java \cite{blanco2020}, software refactoring \cite{peruma2021}, the Internet of Things \cite{uddin2021}, chatbot development \cite{abdellatif2020}, big data \cite{bagherzadeh2019}, and Flutter \cite{alanazi2024}.

Several studies have explored various workplace-related topics, such as interpersonal conflict~\cite{santana2023}, workplace culture~\cite{tassabehji2024problematic}, and emotions and stress~\cite{haque2019managerial,girardi2021emotions}. For example, Santana et al.~\cite{santana2023} analyzed posts from the Software Engineering Stack Exchange website to identify indicators of psychological insecurity in software engineering. They found 11 scenarios of interpersonal conflict, with common issues including divergent ideas, communication breakdowns, poor programming practices, and challenges in time estimation. Newnam and Goode~\cite{newnam2019} analyzed workplace communication patterns in a tech company by examining conversations between supervisors and workers. They categorized communication into task-related, relationship-related, and safety-related types. Their findings showed that over 50\% of conversations were task-related, while relationship and safety-related communication accounted for just 13\%, with the remainder involving third-party discussions or silence. 
Uddin et al.~\cite{uddin2022qualitative} conducted a manual analysis of a small dataset of 825 developer discussions on the DevRant forum during the COVID-19 era. One of the topics they examined was workplace discussions during the pandemic. They found that 75\% of posts addressed themes such as remote work and coworker relationships. Approximately 55\% of these discussions expressed negative sentiment, reflecting pandemic-related challenges, while 32\% were positive, indicating developers' enjoyment of remote work flexibility.

In contrast to the studies mentioned above, our paper employs topic modeling on a large dataset from the Workplace Stack Exchange website to identify various workplace topics, such as interpersonal conflict and communication. None of the existing studies, however, have utilized data from this platform. This paper aims to address this gap by exploring discussions within the Workplace community, providing new insights into developer discussions and participation.

.

\begin{figure}[ht]
     \centering
    {
    \includegraphics[width=\linewidth]{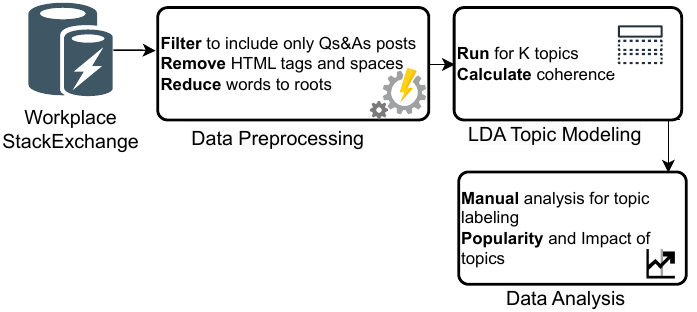}
    }
    \caption{Overview of Methodology}
    \label{fig:method}
\end{figure}

\section{Methodology}

\subsection{Data Collection}
\begin{table}[h]
    \centering
    \caption{Summary of the Workplace Data}
    \begin{tabular}{@{}ll@{}}
        \toprule
        \textbf{Description}                & \textbf{Value} \\ \midrule
        Total \# of posts                   & 143,430        \\
        \#Posts used for LDA (Q and AA)   & 47,368         \\
        Start date                          & 2012-04-10     \\
        End date                            & 2024-05-11     \\
        \#Questions (Q)                   & 32,140         \\
        \#Answers                           & 96,062         \\
                \#Accepted Answers (AA)            & 15,228         \\
        \#Users                             & 107,030\\ \bottomrule
    \end{tabular}
    \label{tab:summary_posts}
\end{table}
The dataset for this study consists of posts from The Workplace Stack Exchange \cite{theWorkplace}, a question-and-answer website dedicated to work-related discussions. It was downloaded from the Stack Exchange Data Explorer \cite{dataExplorer} and included twelve years of posts, from 2012-04-10 to 2024-05-11. Table~\ref{tab:summary_posts} provides a summary of the data. Each post contains the following meta information:
\begin{enumerate}
  \item Post content,
  \item Post creation and edit times,
  \item Post score and view counts,
  \item The user who created the post, and
  \item If it is a question, the tags provided by the user.
\end{enumerate}
An answer is considered accepted when the user who posed the question designates it as the chosen response. Each question can be associated with anywhere from 1 to 5 tags.

The dataset was refined to focus solely on posts containing both questions and their accepted answers, following the approach outlined in previous research~\cite{abdellatif2020,bagherzadeh2019,uddin2021}. This was performed to ensure higher relevance of the answers and to optimize the outcomes of the topic modeling process.

After filtering, the dataset comprised 47,368 posts. In contrast to other studies involving Stack Exchange (e.g.,\cite{bagherzadeh2019, abdellatif2020, alanazi2024, uddin2021, chakraborty2021, santos2023}), this dataset was deliberately not filtered by tags, as the content is already inherently centered around workplace-related topics, making such filtering unnecessary. Fig.~\ref{fig:method} shows the overview of our methodology.

We follow three steps to generate topics from the Workplace posts in the dataset, in line with previous work on topic modeling in Stack Exchange sites~\cite{abdellatif2020,bagherzadeh2019,uddin2021}:



\textit{Step 1: Pre-process the text from the Workplace dataset:}
To prepare the Workplace dataset for analysis, we clean it by removing non-textual elements, such as HTML tags. Next, we filter out irrelevant terms like punctuation marks and symbols. For this, we refer to predefined stop word lists, including those available in libraries like MALLET \cite{mallet} and Python's NLTK \cite{nltk}. Finally, we employ the Porter stemming technique \cite{porter1980} to standardize words to their root forms, which helps consolidate variations like 'organize', 'organized', and 'organizing' into a single base form, 'organ'.

\textit{Step 2: Determine the ideal number of topics}

In our analysis, we leverage Latent Dirichlet Allocation (LDA) \cite{blei2003,mallet} to extract underlying topics from our dataset consisting of workplace-related posts that have undergone preprocessing. To determine the ideal number of topics (\(K\)), we apply a method that identifies the value of \(K\) which yields the highest coherence score, following the procedure outlined by \cite{arun2010}.

Topic coherence is an indicator of the internal consistency of words within each topic. For this evaluation, we use the c\_v metric \cite{rder2015}, which quantifies the relatedness of words in a topic. Through experimentation with various values of \(K\), we observe that a value of \(K = 65\) produces the highest coherence score of 0.5928, which we select as the most appropriate for our modeling.

\textit{Step 3: Generate the final topics}

After determining the optimal number of topics, we generate 65 topics using the dataset and the parameters described. For each topic, we extract the following information:
\begin{itemize}
  \item \textbf{Words}: A list of the top 20 words that best define the topic, along with their associated probabilities.
  \item \textbf{Posts}: A list of posts that are most strongly correlated with each topic, where each post receives a correlation value between 0 and 1. Posts with higher correlation values are more closely aligned with the topic.
\end{itemize}

\subsection{Data Analysis}

\subsubsection{Topic labeling}
\begin{figure*}[t]
    \centering
    {
    \includegraphics[width=\textwidth]{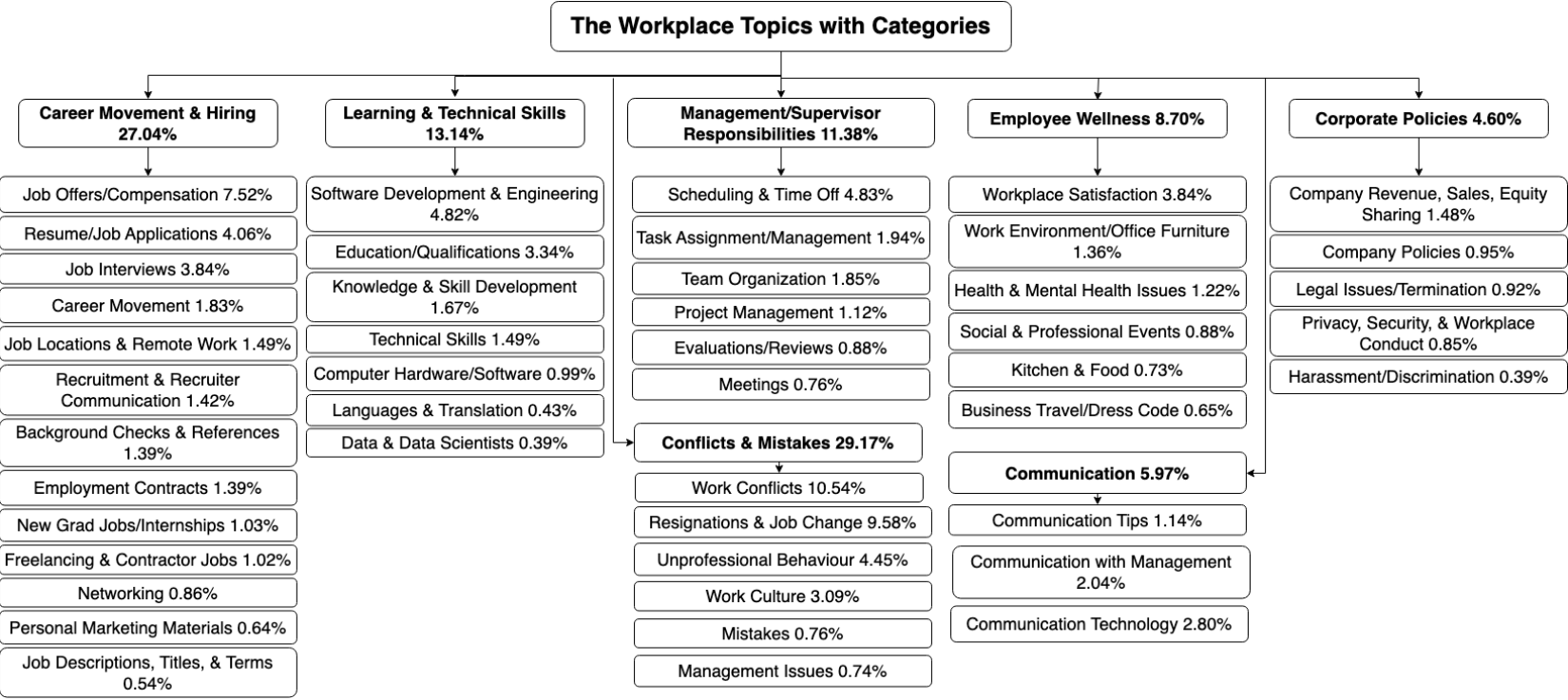}  
    }
    \caption{Topics and Categories}
    \label{fig:topics}
\end{figure*}


Once the number of topics was determined in Step 3, a manual labeling process was carried out for the 65 topics. In line with standard practices \cite{uddin2021,abdellatif2020,bagherzadeh2019}, two researchers were involved in the labeling process to reduce bias. For each topic, a random sample of 20 posts was selected, reviewed, and labeled based on the content and the 20 most frequent words associated with the topic. Any discrepancies between the two researchers were resolved by reviewing additional posts and cross-referencing them with the frequent words. After all 65 topics were labeled, topics with similar labels were merged to eliminate redundancies, resulting in a final set of 46 distinct topics.

\subsubsection{Categorizing topics} 


The 46 labeled topics were then organized into seven categories: Employee Wellness, Communication, Career Movement \& Hiring, Conflicts \& Mistakes, Corporate Policies, Management/Supervisor Responsibilities, and Learning \& Technical Skills. The categorization of the topics was agreed upon by the researchers.

\subsubsection{Topic Absolute and Relative Impact}

To track topic evolution, we computed absolute and relative impacts that were introduced by~\cite{abdellatif2020,bagherzadeh2019,uddin2021}. The LDA algorithm assigns probabilities to each post across topics, with the highest probability indicating the dominant topic. The absolute impact (Eq.~\ref{abs_impact}) is the frequency of a given topic across posts each quarter. Applying LDA to the corpus \( c_j \) yields \( K \) topics \( (z_1, \dots, z_k) \). The absolute impact for topic \( z_k \) in quarter (3 months) \( q \) is:

\begin{equation}
    \text{impact}_{\text{absolute}}(z_k, q) = \sum_{i=1}^{D(q)} I(d_i, z_k)
\label{abs_impact}
\end{equation}

Where \( D(q) \) is the total posts in quarter \( q \), and \( I(d_i, z_k) \):

\begin{equation}
    I(d_i, z_k) =
    \begin{cases}
        1 & \text{if } z_k \text{ is the dominant topic in post } d_i, \\
        0 & \text{otherwise.}
    \end{cases}
\end{equation}

To compute the absolute impact for a category \( C \), which contains multiple topics, we sum the absolute impacts for all topics \( z_k \) within that category. The formula for this is:

\begin{equation}
    \text{impact}_{\text{absolute}}(C, q) = \sum_{z_k \in C} \text{impact}_{\text{absolute}}(z_k, q)
\end{equation}

Where \( C \) is a category (such as "Conflicts \& Mistakes"), and \( z_k \in C \) denotes all topics \( z_k \) that belong to the category \( C \).

To compute the relative impact, we divide the absolute impact by the total number of posts for each topic in a given quarter. The relative impact metric of a topic \( z_k \) in quarter \( q \) is defined in Eq.~\ref{eq:relative_impact}:

\begin{equation}
    \text{impact}_{\text{relative}}(z_k, q) = \frac{1}{D(q)} \sum_{i=1}^{D(q)} I(d_i, z_k)
    \label{eq:relative_impact}
\end{equation}

Where \( D(q) \) is the total number of posts in quarter \( q \) that contain the topic \( z_k \), and \( I(d_i, z_k) \) is the indicator function that equals 1 if the topic \( z_k \) is present in post \( d_i \), and 0 otherwise.

We also compute the relative impact for a category \( C \) in quarter \( q \) by summing the relative impacts of all topics \( z_k \) within that category:

\begin{equation}
    \text{impact}_{\text{relative}}(C, q) = \sum_{z_k \in C} \text{impact}_{\text{relative}}(z_k, q)
\end{equation}

\subsubsection{Topic Popularity and Difficulty}

To assess the popularity of each topic, we calculate two primary metrics: total view count and total score. The view count for posts within a topic typically ranges in the thousands, while the scores are generally between 4 and 23. To make these metrics comparable across topics, we normalize each metric by dividing the topic's value by the total value of that metric across all 46 topics. This yields two new metrics for each topic: one for the normalized view count and one for the normalized score.
Let the normalized view count for topic \(i\) be denoted as \(NormView_i\):

\begin{equation}
    NormView_i = \frac{View_i}{\sum_{j=1}^{46} View_j}
\end{equation}

Similarly, the normalized score for topic \(i\) is represented as \(NormScore_i\):

\begin{equation}
    NormScore_i = \frac{Score_i}{\sum_{j=1}^{46} Score_j}
\end{equation}

To calculate the overall popularity metric, \(OverallPop_i\), for topic \(i\), we compute the average of the normalized view count and score:

\begin{equation}
    FusedPop_i = \frac{NormView_i + NormScore_i}{2}
\end{equation}






We calculate 2 difficulty metrics for each topic: the percentage of questions without an accepted answer and the median time to receive an accepted answer. Again, we normalize the difficulty metrics by dividing the value for each topic by the total value of the metric across all 46 topics. We arrive at two additional metrics for difficulty, one for the normalized percentage of questions without an accepted answer and one for the normalized median time to get an accepted answer.

Let the normalized difficulty metrics for a given topic \(i\) be \(PctNoAccN_i\) and \(MedHrsN_i\):

\begin{equation}
    PctNoAccN_i = \frac{PctNoAcc_i}{\frac{1}{46} \sum_{j=1}^{46} PctNoAcc_j}
\end{equation}

\begin{equation}
    MedHrsN_i = \frac{MedHrs_i}{\frac{1}{46} \sum_{j=1}^{46} MedHrs_j}
\end{equation}

We calculate the fused difficulty (\(FusedD_i\)) of topic \(i\) by taking the average of these two normalized metrics:

\begin{equation}
    FusedDiff_i = \frac{PctNoAccN_i + MedHrsN_i}{2}
\end{equation}

The aforementioned metrics were introduced previously in other StackExchange studies~\cite{bagherzadeh2019,abdellatif2020,uddin2021}.
\section{Results}

\subsection*{RQ1: What workplace topics are discussed on The Workplace Stack Exchange?}

\begin{figure}[t]
    \centering
        \resizebox{\linewidth}{!}{ 
            \includegraphics{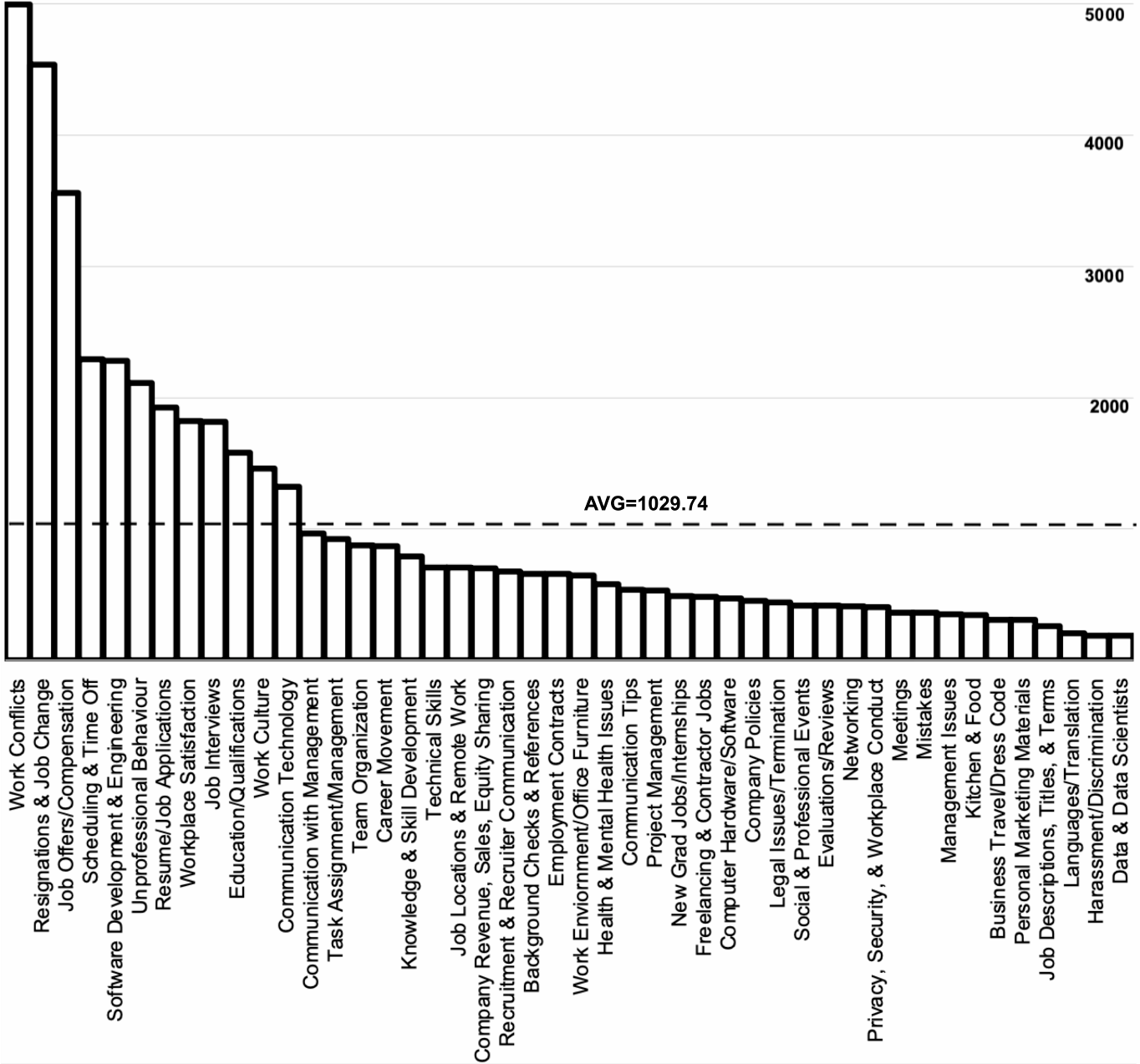}
        }
    
    \caption{Distribution of Topics by Number of Posts}
    \label{fig:topicDistribution}
\end{figure}


\vspace{0.2cm}

After manual analysis of the data, we ended up with 46 topics, divided into 7 categories: Conflict \& Mistakes, Career Movement \& Hiring, Employee Wellness, Learning \& Technical Skills, Manager/Supervisor Responsibilities, Communication, and Corporate Policies. Fig.~\ref{fig:topics} shows the percentage of posts and number of topics for each category. The top two categories are Conflict \& Mistakes, with 29.17\% of posts and 6 topics, and Career Movement \& Hiring with 27.04\% and 13 topics. Although the number of posts between the two categories is similar, Career Movement \& Hiring has significantly more topics than Conflict \& Mistakes. The two categories with the lowest percentage of posts is Communication, 5.97\% of posts and 3 topics, and Corporate Policies, 4.6\% of posts and 5 topics. 



Fig.~\ref{fig:topicDistribution} shows the distribution of posts by topic. The top three topics with the largest number of posts are Work Conflict (4,994), Resignations \& Job Change (4,540), and Job Offers/Compensation (3,560). After those three topics, the number sharply drops, with no more than 2,290 posts per topic. The three topics with the lowest number of posts are Languages/Translation (206), Data \& Data Scientists (186), and Harassment/Discrimination (186). The average number of posts per topic is 1,030.


\subsubsection{Conflict \& Mistakes (29.17\%)} 

This category discusses interpersonal conflict at work between colleagues, management and employees, and how to deal with mistakes made at work. There are 6 topics in this category, and make up the largest percentage of posts by category. Work Conflicts (10.54\%) covers issues related to workplace conflicts, such as handling noisy coworkers (\href{https://workplace.stackexchange.com/questions/64095/how-do-i-deal-with-a-noisy-co-worker}{Q-64095}), or how to handle threats from an employee (\href{https://workplace.stackexchange.com/questions/140633/former-employee-who-thinks-they-were-wronged-is-threatening-me}{Q-140633}). Resignations \& Job Change (9.58\%) include posts on resigning from jobs, changing jobs, recommendation letters, and job fit. \href{https://workplace.stackexchange.com/questions/101281/how-to-explain-that-dont-have-a-recommendation-letter}{Q-101281} and \href{https://workplace.stackexchange.com/questions/23410/poor-performance-review-im-not-a-good-fit}{Q-23410} illustrate this. We chose to include this topic in the Conflict \& Mistakes category with the reasoning that there is often a lot of conflict surrounding resignation between companies and employees, for example \href{https://workplace.stackexchange.com/questions/194540/negotiated-conditions-not-met-after-taking-back-resignation}{Q-194540} and \href{https://workplace.stackexchange.com/questions/171843/can-an-employer-in-the-netherlands-deny-me-holiday-leave-during-my-notice-period}{Q-171843}. Unprofessional Behaviour (4.45\%) includes themes surrounding workplace etiquette, such as how to address a new employee's impolite remarks (\href{https://workplace.stackexchange.com/questions/166232/how-to-address-a-new-employee-that-makes-impolite-and-unprofessional-remarks}{Q-166232}), awkward interactions with coworkers (\href{https://workplace.stackexchange.com/questions/101206/coworker-likes-to-talk-in-the-mens-room}{Q-101206}), and offensive behaviour, including eye rolling, interruptions, and off colour jokes (\href{https://workplace.stackexchange.com/questions/195566/how-to-address-a-coworkers-undermining-behavior-that-affects-my-performance}{Q-195566}). Work Culture (3.09\%) includes posts addressing how companies are structured (\href{https://workplace.stackexchange.com/questions/41737/what-are-the-good-and-bad-effects-of-following-hierarchical-structure-strictly}{Q-41737}), social rules and norms (\href{https://workplace.stackexchange.com/questions/76866/how-can-i-limit-or-avoid-social-faux-pas-when-i-have-a-disability-that-makes-it}{Q-76866}), and office politics (\href{https://workplace.stackexchange.com/questions/30561/why-is-there-a-stigma-in-corporations-against-revealing-hourly-rates-and-salarie}{Q-30561}). Mistakes (0.76\%) involve errors made at work and how to address them. For example, this user asking about what to do when their reputation has been damaged (\href{https://workplace.stackexchange.com/questions/113601/how-do-i-recover-reputation-after-passive-aggressively-challenging-a-management}{Q-113601}), or an intern asking if they should report mistakes they made in project equipment schedules (\href{https://workplace.stackexchange.com/questions/140521/should-i-alert-my-previous-employer-to-errors-i-made-during-an-internship}{Q-140521}). Management Issues (0.74\%) deals with conflict specifically between managers and employees, like disagreeing on productivity standards (\href{https://workplace.stackexchange.com/questions/106953/supervisor-is-saying-productivity-is-low-records-indicate-this-is-not-the-case}{Q-106953}) and assigned workloads (\href{https://workplace.stackexchange.com/questions/113847/coworker-was-fired-manager-expects-the-workload-to-shift-onto-another-team-mem}{Q-113847}).

\subsubsection{Career Movement \& Hiring (27.04\%)}


The Career Movement \& Hiring category is the second largest category, comprising 13 topics. The topics cover everything to do with applying for and finding new jobs. Job Offers/Compensation (7.52\%) include negotiating salary increases (\href{https://workplace.stackexchange.com/questions/178735/how-to-negotiate-a-very-large-salary-increase}{Q-178735}) and accepting job offers (\href{https://workplace.stackexchange.com/questions/100937/accept-a-job-i-dont-want-when-i-know-i-may-get-a-better-offer}{Q-100937}). Resume/Job Applications (4.06\%) covers topics such as the need for a personal statement on a CV (\href{https://workplace.stackexchange.com/questions/98287/do-you-need-a-personal-statement-in-cv}{Q-98287}) and submitting multiple job applications to the same company (\href{https://workplace.stackexchange.com/questions/169724/multiple-job-applications}{Q-169724}). Examples of Job Interview (3.84\%) posts include getting feedback from interviews (\href{https://workplace.stackexchange.com/questions/11537/how-to-get-honest-negative-feedback-from-an-interview}{Q-11537}), and not hearing back from potential employers after interviews (\href{https://workplace.stackexchange.com/questions/168431/not-hearing-back-after-multiple-rounds-of-technical-interviews}{Q-168431}). Career Movement (1.83\%) discusses moving positions within a company (\href{https://workplace.stackexchange.com/questions/102484/want-to-move-to-sales}{Q-102484}), requesting project reassignment (\href{https://workplace.stackexchange.com/questions/106059/how-to-request-project-reassignment-when-handpicked-for-current-position}{Q-106059}), and career development (\href{https://workplace.stackexchange.com/questions/11197/how-can-i-approach-career-development-with-a-boss-who-doesnt-seem-to-support-th}{Q-11197}). Job Location \& Remote Work (1.49\%) include subjects such as working overseas (\href{https://workplace.stackexchange.com/questions/10132/is-long-term-overseas-work-experience-a-merit-or-demerit}{Q-10132}), relocation agreements (\href{https://workplace.stackexchange.com/questions/110648/relocation-agreement}{Q-110648}), and remote working and telecommuting (\href{https://workplace.stackexchange.com/questions/11212/working-for-a-us-company-from-abroad}{Q-11212}). Recruitment \& Recruiter Communication (1.42\%) involves unsolicited calls from recruiters (\href{https://workplace.stackexchange.com/questions/113672/why-do-recruitment-agencies-contact-managers-with-unsolicited-telephone-calls}{Q-113672}), and determining if recruiters are legitimate (\href{https://workplace.stackexchange.com/questions/120925/determining-if-a-recruiter-is-legit}{Q-120925}). Background Checks \& References (1.39\%) include failing background checks (\href{https://workplace.stackexchange.com/questions/119362/failed-background-check-but-not-given-reason}{Q-119362}). Employment Contracts (1.39\%) involve posts that talk about violating a contract by moonlighting (\href{https://workplace.stackexchange.com/questions/14220/how-can-i-minimize-the-risks-of-moonlighting-in-violation-of-my-contract}{Q-14220}), and a contract ending before maternity leave (\href{https://workplace.stackexchange.com/questions/153581/contract-ending-shortly-before-maternal-leave}{Q-153581}). New Grad Jobs/Internships (1.03\%) discuss job offers after internships (\href{https://workplace.stackexchange.com/questions/32251/how-to-handle-a-job-offer-after-an-internship}{Q-32251}) and what employees expect from new grads (\href{https://workplace.stackexchange.com/questions/3425/what-do-employers-expect-from-new-graduates}{Q-3425}). Freelancing \& Contractor Jobs (1.02\%) includes posts about topics such as requesting a reference for a short-term contract (\href{https://workplace.stackexchange.com/questions/112082/asking-for-reference-for-short-contract-job}{Q-112082}) and a company asking an employee to remain silent about freelancing work (\href{https://workplace.stackexchange.com/questions/115430/company-wants-me-to-be-silent-about-doing-freelance-outside-work}{Q-115430}). Networking (0.86\%) consists of posts such as using Linkedin to boost employment prospects (\href{https://workplace.stackexchange.com/questions/1136/how-can-i-use-linkedin-to-maximize-my-employment-prospects}{Q-1136}). Personal Marketing Materials (0.64\%) discusses business cards, personal projects, and portfolios (\href{https://workplace.stackexchange.com/questions/103139/is-it-acceptable-to-include-projects-that-didnt-take-off-in-my-portfolio}{Q-103139}). Job Descriptions, Titles, \& Terms (0.54\%) have to do with the meaning of software job titles (\href{https://workplace.stackexchange.com/questions/107877/meaning-of-senior-in-software-job-titles}{Q-107877}).


\subsubsection{Learning \& Technical Skills (13.14\%)}

\begin{figure}[t]
    \centering
    \includegraphics[width=1\linewidth]{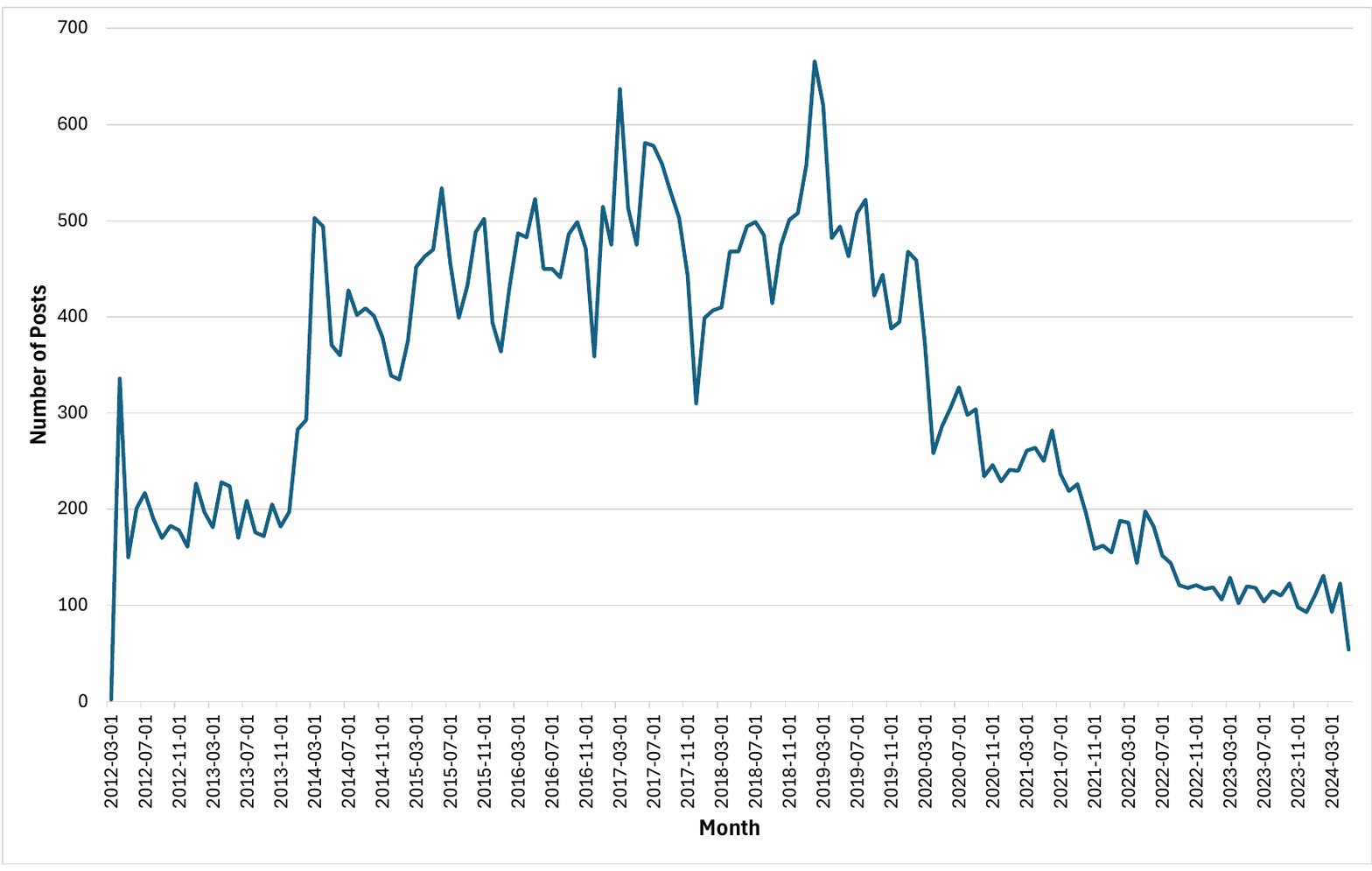}
    \caption{Number of Posts Over Time}
    \label{fig:posts-over-time}
\end{figure}
\begin{figure*}[t]
    \centering
    \includegraphics[width=.65\textwidth]{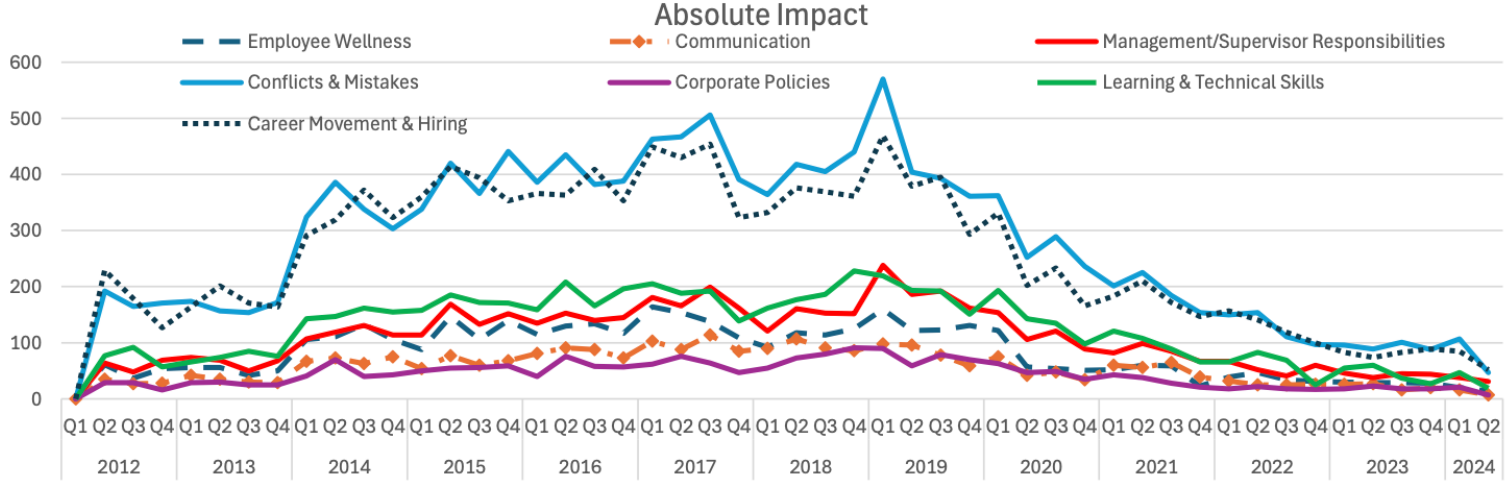}
    \caption{Absolute Impact by Category}
    \label{fig:absolute-impact}
\end{figure*}

\begin{figure*}[t]
    \centering
    \includegraphics[width=.65\linewidth]{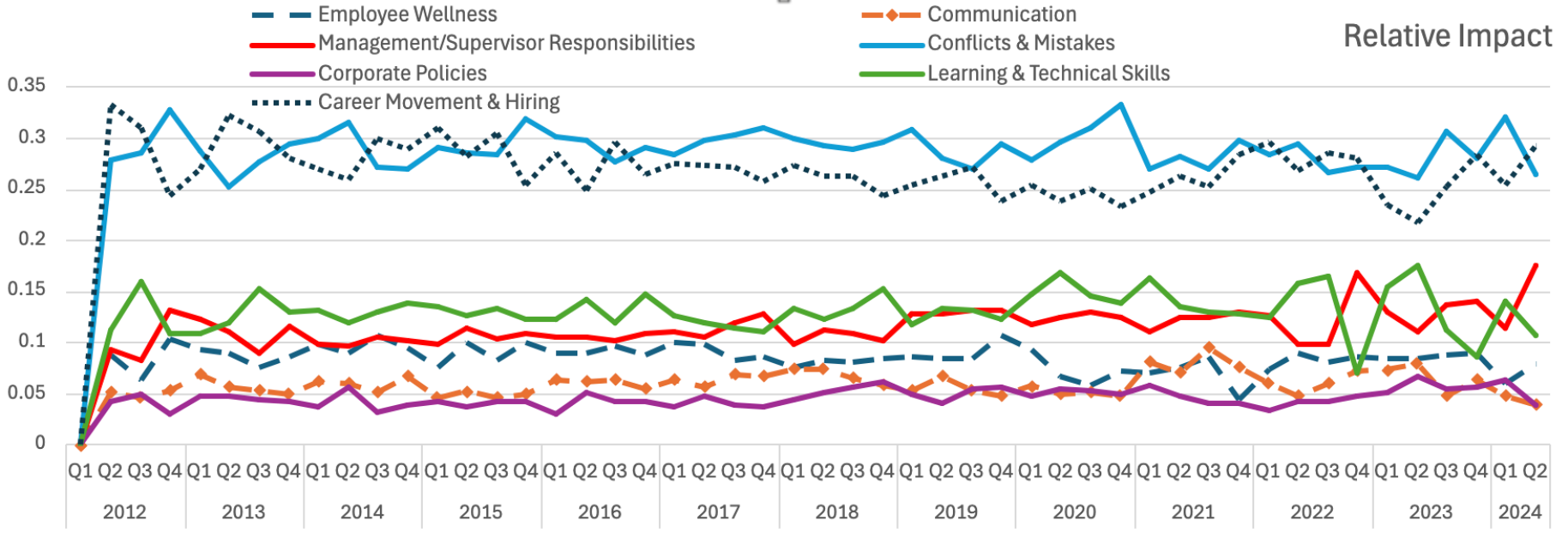}
    \caption{Relative Impact by Category}
    \label{fig:relative-impact}
\end{figure*}
This category contains 7 topics centered on professional growth and skill development. Software Development \& Engineering (4.82\%) is the largest topic in this category. Example posts include questions on code readability (\href{https://workplace.stackexchange.com/questions/102453/code-readability-conventions-and-should-i-let-go}{Q-102453}), when a senior developer suggests changes that break the code (\href{https://workplace.stackexchange.com/questions/115841/after-close-to-two-years-senior-is-still-suggesting-unsolicited-changes-that-br}{Q-115841}), whether or not there is a shortage of software developers (\href{https://workplace.stackexchange.com/questions/124389/is-there-really-a-shortage-of-software-developers}{Q-124389}). Education/Qualifications (3.34\%) consists of discussions about levels of experience(\href{https://workplace.stackexchange.com/questions/50148/how-can-i-properly-describe-my-years-of-experience-on-my-resume}{Q-50148}), and qualifications are needed for certain jobs. One user asks about GPA versus experience (\href{https://workplace.stackexchange.com/questions/112595/what-are-the-advantages-of-a-high-gpa-after-a-few-years-of-experience}{Q-112595}), an another asks whether grades and GPA matters to employers (\href{https://workplace.stackexchange.com/questions/112283/do-employers-look-at-masters-program-grades-and-gpa}{Q-112283}).  Knowledge \& Skill Development (1.67\%) addresses skill development for the workplace, such as a user asking about soft skills and hard skills (\href{https://workplace.stackexchange.com/questions/10302/how-to-avoid-throwing-away-soft-skills-at-a-hard-skills-job}{Q-10302}). Technical Skills (1.49\%) primarily covers coding skills. Computer Hardware/Software (0.99\%) discusses computer tools used in the workplace, e.g. a user asks about using the company laptop for leisure activities (\href{https://workplace.stackexchange.com/questions/104006/is-it-appropriate-to-use-my-company-laptop-for-leisure-activities}{Q-104006}). Languages \& Translation (0.43\%) include posts such as a user asking whether or not they should translate their degree to English on a CV (\href{https://workplace.stackexchange.com/questions/10416/having-a-degree-in-a-non-english-country-should-i-translate-it-to-english-in-th}{Q-10416}), and a user requesting an interview be conducted in English (\href{https://workplace.stackexchange.com/questions/122337/how-to-politely-request-for-the-interview-to-be-conducted-in-english}{Q-122337}). Data \& Data Scientists (0.39\%) covers topics related to data science and data scientist roles, such as a user sharing their experience of being pushed into a new role despite their boss being unhappy with their performance (\href{https://workplace.stackexchange.com/questions/159110/boss-unhappy-with-my-performance-but-is-pushing-me-to-a-new-role-in-big-data-sh}{Q-159110}).



\subsubsection{Management/Supervisor Responsibilities (11.38\%)}

This category consists of 6 topics related to tasks typically handled by managers and supervisors. Scheduling \& Time Off (4.83\%) includes subjects such as holidays and vacation (\href{https://workplace.stackexchange.com/questions/152045/why-are-holidays-stretched-to-the-weekends}{Q-152045}) or work hours (\href{https://workplace.stackexchange.com/questions/106760/how-to-set-boundaries-with-work-hours}{Q-106760}). Task Assignment/Management (1.94\%) relates to scheduling tasks and assignments and time management. Examples include handling unreasonable time estimates from management (\href{https://workplace.stackexchange.com/questions/1633/how-should-reasonable-but-unfavorable-time-estimates-be-handled-by-a-manager}{Q-1633}), and being overloaded with responsibilities (\href{https://workplace.stackexchange.com/questions/34304/working-alone-and-working-overloaded-with-priority-responsibilities-adhd}{Q-34304}). Team Organization (1.85\%) covers team composition and dynamics involved in working within teams. Related posts address situations such as team leaders taking credit for a team member's work (\href{https://workplace.stackexchange.com/questions/10541/how-can-i-deal-with-a-team-lead-who-represents-my-work-as-his-work}{Q-10541}), and ways to motivate team members (\href{https://workplace.stackexchange.com/questions/112660/how-to-handle-and-motivate-a-teammate}{Q-112660}). Project Management (1.12\%) covers topics related to project-based work. Examples include dealing with clients who micromanage (\href{https://workplace.stackexchange.com/questions/146536/client-mis-managing-my-contractor-work}{Q-146536}), and challenges with project scope and accountability (\href{https://workplace.stackexchange.com/questions/150362/being-blamed-for-something-you-dont-consider-your-mistake}{Q-150362}). Meetings (0.76\%) covers various types, including one-on-ones, collaborative, and daily team meetings. One post discusses whether or not skipping meetings affects relationships (\href{https://workplace.stackexchange.com/questions/82729/are-skipping-meetings-at-work-killing-my-interpersonal-relationships}{Q-82729}). Evaluations/Reviews (0.88\%)  covers employee evaluations (\href{https://workplace.stackexchange.com/questions/32144/yearly-employee-evaluation-said-good-things-about-myself-and-a-few-bad-things-a}{Q-32144}) and performance reviews (\href{https://workplace.stackexchange.com/questions/35799/what-to-do-when-your-manager-wont-give-you-your-annual-performance-review}{Q-35799}).

\subsubsection{Employee Wellness (8.70\%)}
This category includes 6 topics focused on employees' mental, physical, and emotional well-being. Workplace Satisfaction (3.84\%) covers discussions such as an employee who is not being challenged at work (\href{https://workplace.stackexchange.com/questions/101517/bored-and-not-challenged-at-good-job}{Q-101517}). Work Environment/Office Furniture (1.36\%) addresses the physical office environment. One user asks if it's appropriate to talk to a rubber duck at the office (\href{https://workplace.stackexchange.com/questions/83451/can-i-talk-to-my-rubber-duck-at-work}{Q-83451}), while others inquire about ergonomic desk setups (\href{https://workplace.stackexchange.com/questions/97644/best-ergonomic-setup-posture-for-people-who-cant-touch-type}{Q-97644}) and wrist pain from their desks (\href{https://workplace.stackexchange.com/questions/72164/desk-is-hurting-my-wrists-what-are-my-options}{Q-72164}). Health \& Mental Health Issues (1.22\%) include posts about sick days, stress leave (\href{https://workplace.stackexchange.com/questions/102626/how-to-correctly-go-on-medical-leave-and-claim-employment-insurance-due-to-workp}{Q-102626}), illness, depression (\href{https://workplace.stackexchange.com/questions/119518/should-i-be-talking-to-hr-about-my-struggles-with-stress-and-depression-at-my-wo}{Q-119518}), dental benefits, and medication (\href{https://workplace.stackexchange.com/questions/124032/does-being-prescribed-anti-depressants-look-bad-to-employers-during-security-cl}{Q-124032}). Social \& Professional Events (0.88\%) include social work events such as work Christmas parties (\href{https://workplace.stackexchange.com/questions/188956/is-it-legal-for-an-employer-in-fl-to-tell-employees-who-agreed-to-attend-the-com}{Q-188956}) and team building activities (\href{https://workplace.stackexchange.com/questions/3105/team-building-activities-for-a-fast-growing-team}{Q-3105}). Kitchen \& Food (0.73\%) covers topics related to work kitchens, lunches, and food. Examples include posts about food theft from shared fridges (\href{https://workplace.stackexchange.com/questions/6283/how-can-we-prevent-theft-of-food-from-shared-fridges-by-coworkers}{Q-6283}) and shared coffee mugs (\href{https://workplace.stackexchange.com/questions/36368/a-co-worker-is-using-my-cup}{Q-36368}). Business Travel/Dress Code (0.65\%) covers appropriate work clothing (\href{https://workplace.stackexchange.com/questions/66181/suitable-company-wear}{Q-66181}) and uniforms (\href{https://workplace.stackexchange.com/questions/60097/should-an-employer-be-required-to-provide-well-fitting-uniforms}{Q-60097}).


\subsubsection{Communication (5.97\%)}

Fatimayan \cite{fatimayin2018} defines communication as an interaction within a social context involving a sender and a receiver. In this case, we refer to communication in the workplace. The Communication category consists of three topics. Communication Technology (2.8\%) covers communication devices like email, phones, and tools such as Slack or Zoom. Example posts include questions about email signatures (\href{https://workplace.stackexchange.com/questions/151328/can-i-put-my-business-signature-on-my-personal-email}{Q-151328}) and coworkers overusing video conferencing (\href{https://workplace.stackexchange.com/questions/190621/how-to-deal-with-a-coworker-who-only-wants-to-communicate-via-video-conference}{Q-190621}). Communication with Management (2.04\%) pertains to conversations between employees and management. Topics include promotion requests (\href{https://workplace.stackexchange.com/questions/191650/handling-a-managers-irritating-behaviour-rejecting-promotion-request}{Q-191650}). Communication Tips (1.14\%) refers to posts seeking advice on how to communicate and discuss difficult issues. Examples include communicating with customers (\href{https://workplace.stackexchange.com/questions/105857/telling-the-customer-it-has-lower-priority}{Q-105857}).

\subsubsection{Corporate Policies (4.6\%)}

This category consists of 5 topics. Company Revenue, Sales, Equity Sharing (1.48\%) consists of posts about company equity (\href{https://workplace.stackexchange.com/questions/96384/is-it-common-to-find-software-jobs-that-offer-part-ownership-in-company}{Q-96384}) and retirement benefits (\href{https://workplace.stackexchange.com/questions/97643/what-does-a-company-gain-by-allowing-an-unusually-early-full-401k-vesting}{Q-97643}). Company Policies (0.95\%) addresses issues such as pay policies (\href{https://workplace.stackexchange.com/questions/17684/employer-pays-wages-2-weeks-late-what-if-anything-to-do-about-it}{Q-17684}) and drug tests (\href{https://workplace.stackexchange.com/questions/159029/is-it-possible-to-be-drug-tested-by-a-ca-employer-while-working-from-home}{Q-159029}). Legal Issues/Termination (0.92\%) includes posts relating to cause for termination (\href{https://workplace.stackexchange.com/questions/104298/when-can-someone-be-fired-for-just-cause}{Q-104298}), employee theft (\href{https://workplace.stackexchange.com/questions/122628/how-should-i-deal-with-an-employee-who-is-stealing-from-the-cash-counter}{Q-122628}), written warnings, layoffs and severance pay (\href{https://workplace.stackexchange.com/questions/40727/questions-about-layoffs}{Q-40727}). Privacy, Security, and Workplace Conduct (0.85\%) refers to information security of employees, clients, and confidentiality. Posts include subjects such as email phishing (\href{https://workplace.stackexchange.com/questions/159349/employee-test-result-failure-revealed}{Q-159349}), information leaks (\href{https://workplace.stackexchange.com/questions/54999/leaked-confidential-information-about-a-management-decision-should-i-make-the-m}{Q-54999}), and employee confidentiality (\href{https://workplace.stackexchange.com/questions/75243/can-a-company-have-access-to-an-employees-counselling-session-notes}{Q-75243}). Harassment/Discrimination (0.39\%) includes posts such as sexual harassment in the workplace (\href{https://workplace.stackexchange.com/questions/104314/what-happens-when-hr-is-the-source-of-harassment}{Q-104314}) and racism (\href{https://workplace.stackexchange.com/questions/138741/how-to-deal-with-a-manager-who-frequently-makes-subtle-racist-remarks-about-my}{Q-138741}).

\addtocounter{o}{1}
\begin{tcolorbox}[flushleft
upper,boxrule=1pt,arc=0pt,left=0pt,right=0pt,top=0pt,bottom=0pt,colback=white,after=\ignorespacesafterend\par\noindent]
Stack Exchange's Workplace covers 46 topics, with "Work Conflicts" having the most posts, followed by "Resignations \& Job Change". These topics fall into seven main categories: Employee Wellness, Communication, Career Movement \& Hiring, Conflicts \& Mistakes, Corporate Policies, Management/Supervisor Responsibilities, and Learning \& Technical Skills.

\end{tcolorbox}

 \subsection*{RQ2: How do the topics evolve over time?}

We analyzed topic trends over time by examining the number of posts per quarter. Fig.~\ref{fig:posts-over-time} illustrates the total posts over time, with an initial spike in activity when The Workplace launched in March 2013. For the remainder of 2013, posts per month stayed around 200, increasing to 500 in early 2014. Posting was at its highest from March 2014 to March 2020, peaking in March 2019, followed by a steady decline. This decline accelerated after the COVID-19 period, with monthly posts falling below 200 from November 2021 onward, reaching their lowest levels in 2023 and 2024.
 

We further analyzed how each category evolves over time by tracking its impact. One way to measure this is by calculating the absolute impact, which involves tracking the number of new posts added each month, as outlined in Section 3.

We observe that the absolute impact across all categories follows a similar trend, peaking between 2014 and 2020, and then declining to its lowest point between 2022 and 2024, as shown in Fig. \ref{fig:absolute-impact}. Consistent with the results from RQ1, Conflicts \& Mistakes has the highest absolute impact, followed by Career Movement \& Hiring. The impact of these two categories peaks around 2019. This could be because the tech sector faced increasing turnover rates, a report from the Work Institute indicated that 27\% of U.S. workers left voluntarily in 2018, with this trend continuing in subsequent years~\cite{workinstitute2019,aura2019}. After this peak, the trends for these categories converge with those of the rest, particularly following the COVID-19 period. The category with the lowest absolute impact is Corporate Policies.

We also measure the relative impact, which refers to the number of new posts added to a category each month relative to other categories, as explained in Section 3. Similar to absolute impact, Conflicts \& Mistakes has the highest relative impact, with some overlap with Career Movement \& Hiring, as shown in Fig. \ref{fig:relative-impact}. Career Movement \& Hiring surpasses Conflicts \& Mistakes mainly between 2022 and 2024, and again between 2012 and 2015. The relative impact for Career Movement \& Hiring spikes in March 2024. Career movement is a major concern for developers, with many dissatisfied by limited career advancement, driving them to seek new roles or employers. The lack of growth opportunities significantly contributes to high turnover rates in tech industry~\cite{workinstitute2019,stackoverflow2019}.
Overall, the relative impact for all categories remains relatively stable. The category with the lowest overall relative impact is Corporate Policies.

\addtocounter{o}{1}
\begin{tcolorbox}[flushleft
upper,boxrule=1pt,arc=0pt,left=0pt,right=0pt,top=0pt,bottom=0pt,colback=white,after=\ignorespacesafterend\par\noindent]
Topics vary in terms of absolute and relative impact. We found that the "Conflicts \& Mistakes" category and "Career Movement \& Hiring" have the highest absolute and relative impact. Additionally, we found that topics in the "Corporate Policies" category have the lowest absolute and relative impact.

\end{tcolorbox}


\begin{figure}
    \centering
    \includegraphics[width=1\linewidth]{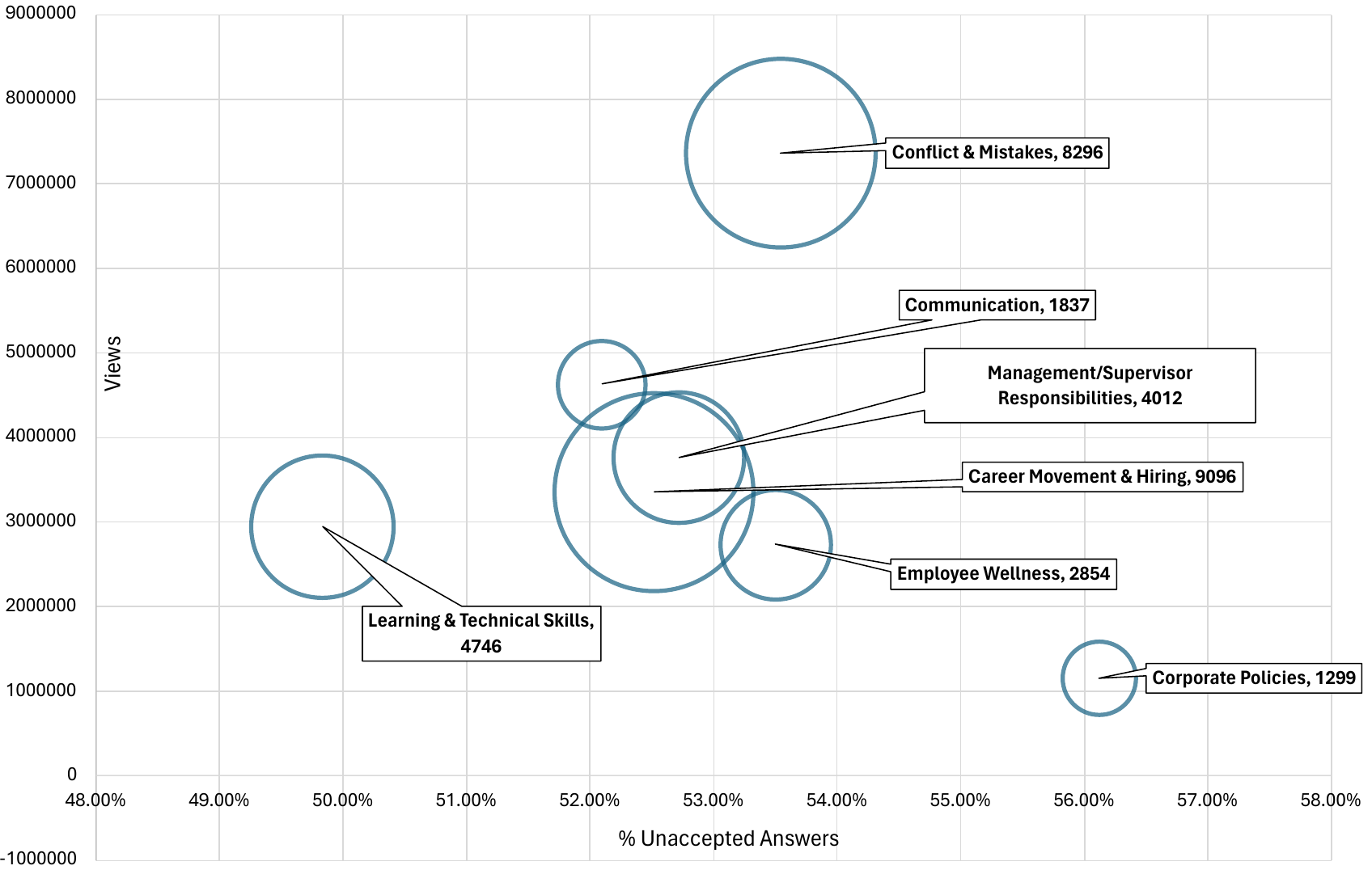}
    \caption{Difficulty and Popularity of Categories}
    \label{fig:scatter}
\end{figure}

\begin{table}[ht]
\centering
\scriptsize
\setlength{\tabcolsep}{1.5pt} 
\renewcommand{\arraystretch}{1.1} 
\caption{Popularity of Topics} 
\label{tab:topic_popularity} 
\begin{tabular}{|l|l|c|c|c|c|c|c|}
\hline
Topic Label & Category & Poplrty & Qs & \#Views & \#Score & A.Views & A.Score \\
\hline
Work Conflicts & Conflicts & 5.59 & 3384 & 18256151 & 44924 & 5395 & 13.3 \\
Job Offers & Career Movmnt & 3.42 & 2559 & 13934973 & 21976 & 5446 & 8.6 \\
Scheduling & Management & 3.41 & 1795 & 13393383 & 22883 & 7462 & 12.7 \\
Resignations & Conflicts & 3.33 & 2470 & 13713505 & 21013 & 5552 & 8.5 \\
Software Dev & Learning & 2.49 & 1807 & 7104403 & 22056 & 3932 & 12.2 \\
Unprofessional Behv & Conflicts & 2.46 & 1322 & 7538877 & 20709 & 5703 & 15.7 \\
Job Interviews & Career Movmnt & 2.19 & 1430 & 8878922 & 14129 & 6209 & 9.9 \\
Comm Tech & Communication & 1.83 & 876 & 9511758 & 7648 & 10858 & 8.7 \\
Wrkplce Satisfaction & Wellness & 1.57 & 1159 & 5102377 & 12718 & 4402 & 11 \\
Resume & Career Movmnt & 1.22 & 1164 & 5262438 & 7255 & 4521 & 6.2 \\
Qualifications & Learning & 1.03 & 1286 & 4377682 & 6293 & 3404 & 4.9 \\
Health & Wellness & 0.88 & 453 & 3087178 & 6689 & 6815 & 14.8 \\
Work Culture & Conflicts & 0.87 & 662 & 2544024 & 7567 & 3843 & 11.4 \\
Work Environment & Wellness & 0.86 & 483 & 2794273 & 6949 & 5785 & 14.4 \\
Task Assignment & Management & 0.86 & 588 & 2753593 & 6989 & 4683 & 11.9 \\
Team Organization & Management & 0.84 & 711 & 2476997 & 7307 & 3484 & 10.3 \\
Comm w/Mmgmntt & Communication & 0.79 & 630 & 2718598 & 6037 & 4315 & 9.6 \\
Hardware & Learning & 0.67 & 348 & 2018616 & 5684 & 5801 & 16.3 \\
Background Checks & Career Movmnt & 0.63 & 487 & 3239262 & 2720 & 6652 & 5.6 \\
Skill Development & Learning & 0.63 & 412 & 2195238 & 4737 & 5328 & 11.5 \\
Comm. Tips & Communication & 0.59 & 331 & 1638809 & 5385 & 4951 & 16.3 \\
Kitchen\&Food & Wellness & 0.56 & 255 & 1707695 & 4797 & 6697 & 18.8 \\
Technical Skills & Learning & 0.56 & 607 & 2018502 & 4150 & 3325 & 6.8 \\
Job Locations & Career Movmnt & 0.56 & 568 & 1999232 & 4186 & 3520 & 7.4 \\
Business Travel & Wellness & 0.53 & 219 & 1977209 & 3800 & 9028 & 17.4 \\
Social Events & Wellness & 0.52 & 285 & 1727964 & 4134 & 6063 & 14.5 \\
Company Revenue & Corporate Policy & 0.50 & 512 & 1666179 & 4019 & 3254 & 7.8 \\
Languages & Learning & 0.49 & 140 & 2493204 & 2129 & 17809 & 15.2 \\
Career Movement & Career Movmnt & 0.48 & 601 & 1491230 & 4040 & 2481 & 6.7 \\
Project Mngmnt & Management & 0.48 & 435 & 1515928 & 3896 & 3485 & 9 \\
Empl. Contract & Career Movmnt & 0.45 & 371 & 1566194 & 3424 & 4222 & 9.2 \\
Recruitment & Career Movmnt & 0.45 & 475 & 1712814 & 3096 & 3606 & 6.5 \\
Networking & Career Movmnt & 0.42 & 295 & 1259926 & 3609 & 4271 & 12.2 \\
Legal Issues & Corporate Policy & 0.42 & 227 & 1240693 & 3616 & 5466 & 15.9 \\
Harassment & Corporate Policy & 0.37 & 142 & 1025580 & 3316 & 7222 & 23.4 \\
Meetings & Management & 0.36 & 218 & 1180206 & 2891 & 5414 & 13.3 \\
Freelancing & Career Movmnt & 0.36 & 358 & 1241060 & 2747 & 3467 & 7.7 \\
Evaluations & Management & 0.34 & 265 & 1245199 & 2521 & 4699 & 9.5 \\
Marketing Materials & Career Movmnt & 0.33 & 218 & 1252123 & 2247 & 5744 & 10.3 \\
Mistakes & Conflicts & 0.32 & 165 & 1124874 & 2418 & 6817 & 14.7 \\
Management Issues & Conflicts & 0.27 & 293 & 997569 & 1984 & 3405 & 6.8 \\
New Grad Jobs & Career Movmnt & 0.26 & 407 & 1071359 & 1612 & 2632 & 4 \\
Company Policy & Corporate Policy & 0.26 & 209 & 1024167 & 1693 & 4900 & 8.1 \\
Privacy & Corporate Policy & 0.25 & 209 & 818274 & 1977 & 3915 & 9.5 \\
Job Descriptions & Career Movmnt & 0.16 & 163 & 681328 & 984 & 4180 & 6 \\
Data Science & Learning & 0.13 & 146 & 421255 & 1040 & 2885 & 7.1 \\
\hline
\end{tabular}
\end{table}

\begin{table}[ht]
\centering
\scriptsize
\setlength{\tabcolsep}{1.5pt} 
\renewcommand{\arraystretch}{1.1} 
\caption{Difficulty of Topics} 
\label{tab:topic_difficulty} 
\begin{tabular}{|l|l|c|c|c|c|c|c|}
\hline
\textbf{Topic Label} & \textbf{Category} & \textbf{ Difficulty} & \textbf{No AA} & \textbf{Time} \\
\hline
Data Science & Learning  & 1.44 & 0.48 & 1.88 \\
Work Culture & Conflicts & 1.36 & 0.51 & 1.66 \\
Harassment/Discrimination & Corporate Policies & 1.27 & 0.48 & 1.55 \\
Company Policies & Corporate Policies & 1.27 & 0.55 & 1.42 \\
Company Revenue & Corporate Policies & 1.20 & 0.58 & 1.23 \\
Team Organization & Management & 1.18 & 0.49 & 1.35 \\
Health \& Mental Health Issues & Employee Wellness & 1.16 & 0.59 & 1.13 \\
Communication with Management & Communication & 1.15 & 0.59 & 1.12 \\
Career Movement & Career Movement & 1.15 & 0.57 & 1.15 \\
Knowledge \& Skill Development & Learning  & 1.13 & 0.5 & 1.24 \\
Communication Tips & Communication & 1.09 & 0.52 & 1.13 \\
Job Locations \& Remote Work & Career Movement & 1.08 & 0.53 & 1.1 \\
 Workplace Conduct & Corporate Policies & 1.05 & 0.51 & 1.08 \\
Legal Issues/Termination & Corporate Policies & 1.05 & 0.63 & 0.85 \\
Management Issues & Conflicts & 1.05 & 0.61 & 0.88 \\
Education/Qualifications & Learning  & 1.04 & 0.54 & 1 \\
Software Dev & Learning  & 1.02 & 0.48 & 1.07 \\
Project Management & Management & 1.01 & 0.53 & 0.95 \\
Workplace Satisfaction & Employee Wellness & 0.98 & 0.55 & 0.87 \\
Technical Skills & Learning  & 0.97 & 0.48 & 0.98 \\
Computer Hardware/Software & Learning  & 0.97 & 0.52 & 0.9 \\
Freelancing \& Contractor Jobs & Career Movement & 0.97 & 0.51 & 0.91 \\
Mistakes & Conflicts & 0.97 & 0.45 & 1.02 \\
Unprofessional Behaviour & Conflicts & 0.97 & 0.52 & 0.89 \\
Task Assignment/Management & Management & 0.96 & 0.54 & 0.85 \\
Work Conflicts & Conflicts & 0.96 & 0.52 & 0.88 \\
Job Offers/Compensation & Career Movement & 0.96 & 0.55 & 0.82 \\
Employment Contracts & Career Movement & 0.95 & 0.53 & 0.85 \\
Personal Marketing Materials & Career Movement & 0.95 & 0.48 & 0.93 \\
Resignations \& Job Change & Conflicts & 0.94 & 0.56 & 0.77 \\
Scheduling \& Time Off & Management & 0.94 & 0.55 & 0.78 \\
Background Checks & Career Movement  & 0.93 & 0.63 & 0.62 \\
Evaluations/Reviews & Management & 0.93 & 0.5 & 0.86 \\
Work Environment/ Furniture & Employee Wellness & 0.92 & 0.53 & 0.78 \\
Meetings & Management & 0.92 & 0.49 & 0.85 \\
Resume/Job Applications & Career Movement  & 0.89 & 0.49 & 0.8 \\
Job Descriptions & Career Movement  & 0.89 & 0.53 & 0.72 \\
Kitchen \& Food & Employee Wellness & 0.86 & 0.51 & 0.7 \\
Recruitment  & Career Movement  & 0.85 & 0.52 & 0.67 \\
Business Travel/Dress Code & Employee Wellness & 0.84 & 0.48 & 0.72 \\
Communication Technology & Communication & 0.83 & 0.47 & 0.72 \\
Social \& Professional Events & Employee Wellness & 0.82 & 0.46 & 0.73 \\
Job Interviews & Career Movement  & 0.81 & 0.49 & 0.65 \\
Networking & Career Movement  & 0.80 & 0.46 & 0.69 \\
New Grad Jobs/Internships & Career Movement  & 0.80 & 0.47 & 0.67 \\
Languages/Translation & Learning  & 0.74 & 0.4 & 0.68 \\
\hline
\end{tabular}
\end{table}
 \subsection*{RQ3: What is the popularity and difficulty of topics?}

Popularity and difficulty metrics provide valuable insights into research efforts, helping to assess the focus and impact of different topics. These metrics highlight areas gaining traction and those that require further investigation to address emerging challenges~\cite{abdellatif2020,uddin2021,bagherzadeh2019}. Table~\ref{tab:topic_popularity} presents the popularity of each topic, calculated using the metrics described in Section 3.

Work Conflicts is the most popular topic with a fused popularity of 5.59. Of the top five most popular topics, it has the highest number of questions, total views, and total score. Languages has the highest number of average views per question at 17,809, whereas Work Conflicts has 5,395.

Fused difficulty is determined by the percentage of questions without an accepted answer and the median time it takes for a post to receive an accepted answer. Table~\ref{tab:topic_difficulty} presents the difficulty of each topic. Three of the five most difficult topics belong to the Corporate Policies category: Harassment/Discrimination, Company Policies, and Company Revenue, Sales, Equity Sharing. Topics from Company Policies also rank among the five least popular topics, suggesting that questions in the Corporate Policies category are both difficult to answer and unpopular. This aligns with findings from studies indicating that corporate policies are often seen as burdensome and confusing by developers. For instance, Vassallo et al. noted that policies related to security, privacy, and compliance can hinder development workflows, leading to frustration and delays~\cite{vassallo2020corporate}. Additionally, Feller et al. emphasized the challenges developers face when organizational policies are not flexible enough to support agile practices~\cite{feller2020adapting}.

Interestingly, the Data \& Data Scientists category is the least popular of all 46 topics and is also the most difficult, with only 40\% of questions receiving accepted answers and a median time of 1.88 hours for an answer to be accepted. Three topics from the Career Movement \& Hiring category rank among the least difficult topics: Job Interviews, Networking, and New Grad Jobs/Internships. All three topics have between 46-49\% of questions without accepted answers and between 0.65-0.69 hours for an answer to be accepted. Languages/Translation is the least difficult topic.

We wanted to explore the potential inverse relationship between topic popularity and difficulty. For example, \textit{Data Science} is both the least popular and the most difficult. However, this trend is less clear for Work Culture, which ranks second in difficulty based on unanswered questions but 13th in popularity. To quantitatively assess the correlation between popularity and difficulty, we used the Kendall Tau correlation measure~\cite{Kendall-TauMetric-Biometrica1938}, which is less sensitive to outliers compared to the Mann-Whitney correlation~\cite{Kruskal-Wilcoxon-JASS1957}.

Table~\ref{tab:correlation} presents the correlations between two popularity metrics (Scores and Views) and two difficulty metrics (percentage of unanswered questions and median hours to accepted answers). The results show no strong statistical evidence of a correlation between difficulty and popularity, indicating that popular topics are not necessarily more difficult.

Fig.~\ref{fig:scatter} illustrates the difficulty and popularity of each category. The x-axis represents the percentage of answers that remain unaccepted, while the y-axis reflects the number of views. As shown in the figure, Corporate Policies emerges as the most challenging category, yet it attracts the fewest views. In contrast, Conflicts \& Mistakes garners the highest number of views despite being moderately difficult. Meanwhile, Learning \& Technical Skills ranks as the least challenging category overall.

\addtocounter{o}{1}
\begin{tcolorbox}[flushleft
upper,boxrule=1pt,arc=0pt,left=0pt,right=0pt,top=0pt,bottom=0pt,colback=white,after=\ignorespacesafterend\par\noindent]
Topics vary in terms of popularity and difficulty. We found that "Work Conflict" is the most popular topic, and the "Conflicts \& Mistakes" category has the highest number of views overall. Additionally, we found that topics in the "Corporate Policies" category are the most difficult, with about 56\% of the questions in this category lacking accepted answers.
\end{tcolorbox}

\begin{table}[ht]
\centering
\scriptsize
\setlength{\tabcolsep}{10pt} 

\renewcommand{\arraystretch}{1.3} 
\caption{Coefficients and p-values for Views and Scores}
\begin{tabular}{|l|l|l|}
\hline
\textbf{Coefficient/p-value} & \textbf{Views} & \textbf{Scores} \\
\hline
\textbf{w/o Accepted Answer} & 0.090/0.379 & 0.090/0.379 \\
\hline
\textbf{Median Hours to acc. Answer} & (-0.151)/0.140 & (-0.002)/0.985 \\
\hline
\end{tabular}
\label{tab:correlation}
\end{table}

\begin{table}[h!]
\centering
\begin{tabular}{|l|l|}
\hline
\textbf{Posts} & \textbf{Users} \\
\hline
0 & 77,935\\
1-10 & 27,532\\
11-50 & 1211 \\
51-100 & 172 \\
101-200 & 100 \\
201-300 & 26 \\
301-400 & 15 \\
401-500 & 10 \\
501-1000 & 19 \\
$>$1000 & 10 \\
\hline
\end{tabular}
\caption{Distribution of Users by Number of Posts}
\label{tab:posts-users}
\end{table}

\begin{table}[ht]
\centering
\begin{tabular}{|c|c|c|c|}
\hline
\textbf{Category} & \textbf{Precision} & \textbf{Recall} & \textbf{F1-score} \\
\hline
Men & 0.88 & 1.00 & 0.936 \\
Women & 0.83 & 0.99 & 0.902 \\
\hline

\hline
\end{tabular}
\caption{Precision, Recall, and F1-scores for Men and Women}
\label{tab:precision-reacll}
\end{table}

\begin{table}[h!]
\centering
\begin{tabular}{|l|c|c|c|c|}
\hline
\textbf{Categories} & \textbf{\#Women} & \textbf{\%Women} & \textbf{\#Men} & \textbf{\%Men} \\
\hline
Employee Wellness & 507 & 11\% & 2665 & 10\% \\
Communication & 299 & 7\% & 1774 & 6\% \\
Career Movement  & 1159 & 25\% & 7227 & 26\% \\
Management & 636 & 14\% & 3534 & 13\% \\
Conflicts\&Mistakes & 1155 & 25\% & 6808 & 24\% \\
Corporate Policies & 207 & 5\% & 1286 & 5\% \\
Learning  & 624 & 14\% & 4507 & 16\% \\
\hline
\textbf{Total by Gender} & 4587 & & 27,801& \\
\hline
\end{tabular}
\caption{Posts by category and gender}
\label{tab:gender-posts}
\end{table}


\subsection*{RQ4: How does user activity and gender distribution vary across The Workplace discussion community?}

We aimed to understand the frequency of posts per user and explore the dynamics of community participation by grouping users into posting brackets (e.g., fewer than 10 posts, 10–50 posts, etc.). This approach reveals patterns of engagement across a range of activity levels, allowing us to identify “superusers” or dominant participants who contribute disproportionately to discussions. Research on online community dynamics shows that superusers can significantly shape interactions, driving both positive and negative effects on engagement and community inclusiveness~\cite{huang2018power,kumar2019influence}. 

Table~\ref{tab:posts-users} shows the distribution of posts across users. We see that the overwhelming majority of Workplace users have no posts, suggesting they may be merely viewers who are interested in the discussions. A very large number of users have between 1-10 posts. Only 10 users have more than 1,000 posts. The top user has 3,942 posts, which constitutes 2.7\% of the total dataset. The top 20 users, ranked by the number of posts, have contributed 24,689 posts, or 17.2\% of the total dataset. The top user's account was created in 2013. Most of the posts by this user are in \textit{Career Movement \& Hiring} (1,274 posts) and \textit{Conflicts \& Mistakes} (1,020 posts). The fewest posts by this user are in \textit{Corporate Policies} (164 posts). Among the top 20 users by reputation, the majority of their posts are also in \textit{Career Movement \& Hiring} (6,578 posts) and \textit{Conflicts \& Mistakes} (6,638 posts). The lowest number of posts is in \textit{Corporate Policies} (1,075 posts), followed by \textit{Communication} (1,460 posts).


Additionally, we analyzed the gender distribution across topics. To determine a user's gender, we processed a dataset of posts with users' display names through genderize.io~\cite{genderizeio, noei2022study}, using only results with over 90\% certainty for a particular gender and labeling the remaining users as unknown. Out of 107,030 users on the Workplace site, genderize.io was able to identify 4,587 women and 27,801 men. About 14\% of identified users were women and 86\% were men. A large proportion of users either opt not to use their real names or use pseudonyms instead. Since we used an automated approach for gender assignment, we wanted to manually validate the results. We selected a statistically significant sample of 383 users, ensuring a 95\% confidence level with a 5\% margin of error. Following related research~\cite{reid2010role}, we searched for the users' names online and examined photos of the users whose gender was unclear. We then calculated the precision and recall for each race/gender group. Precision calculates the proportion of identified positive class instances that truly belong to the positive class, i.e., Precision $=\frac{True Positives}{True Positives + False Positives}$. Recall calculates the proportion of identified positive class instances from all positive class instances in the dataset, i.e., Recall $=\frac{True Positives}{True Positives + False Negatives}$. F-measure is a metric that balances both precision and recall. Table~\ref{tab:precision-reacll} reports the precision and recall for men and women. Overall, the precision and recall for both groups are quite high.

We then analyzed the posts by gender and category. Table~\ref{tab:gender-posts} shows the detailed breakdown of posts per category for both women and men users. For each gender, we calculated the percentage of posts in each category relative to the total number of posts by that gender. The percentages for women and men were very similar. However, the number of posts per category based on gender was considerably different. For instance, women users posted 25\% of their overall posts in the "Conflicts \& Mistakes" category, while men posted 24\%. The total number of posts in this category for women was 1,155, whereas men  posted 6,808. In the "Corporate Policies" category, both women and men contributed 5\% of their total posts, with women making 207 posts and men making 1,286. This analysis shows that while women and men post similar percentages across categories, men contribute significantly more posts overall. This disparity reflects a gender gap in total participation, despite proportional parity in engagement. We conducted a Chi-Square test for independence to examine the relationship between gender and post categories. The null hypothesis assumed no association between gender and post categories. The results indicated a significant association, $\chi^2(6) = 31.42$, $p = 0.000021$. This suggests that gender significantly influences the distribution of posts across different categories, leading us to reject the null hypothesis.

\addtocounter{o}{1}
\begin{tcolorbox}[flushleft
upper,boxrule=1pt,arc=0pt,left=0pt,right=0pt,top=0pt,bottom=0pt,colback=white,after=\ignorespacesafterend\par\noindent]
A small number of users tend to dominate the discussions. We found that the top 20 users contribute to 17.2\% of the posts, while the top user contributes to 2.7\% of the posts. We also found that there are more men contributing to discussions than women (14\% women and 86\% men). While men and women post similar proportions across categories, we found a significant association between gender and post categories.

\end{tcolorbox}

\section{Threats to Validity}

\textbf{External Validity:} The generalizability of our findings is limited by the specific context of this study. We focus on Workplace StackExchange, one of the largest and most popular Q\&A websites for workplace discussions among developers. Our findings may not generalize to developers who are not active in online forums. Previous studies, such as Zhang et al.~\cite{zhang2019study} and Xie et al.~\cite{xie2020understanding}, have noted that Q\&A platforms in other domains may exhibit distinct user behavior patterns and content dynamics. These differences may affect how questions are posed and answered in those contexts. Therefore, further research is needed to determine whether similar trends exist among developers who are not involved in Q\&A platforms.

\textbf{ Internal Validity } Potential biases and analysis errors pose a threat to the validity of our findings, particularly due to the manual labeling and categorization of topics involved in our study. This process inherently carries a risk of subjective influence. To mitigate this, we incorporated two independent raters to conduct the analysis, fostering a more balanced and impartial approach. Using multiple raters aligns with established qualitative research practices, which emphasize reducing individual biases and enhancing the reliability of findings~\cite{Saldana2016}.

\textbf{Construct Validity} Threats related to the difficulty in finding data relevant to workplace discussions are notable. We collected all posts from the Workplace StackExchange site, avoiding the use of tags, which previous research relied on~\cite{uddin2021,tahir2020,chakraborty2021}, as our dataset from the site is homogeneous. For the gender analysis, we used the users table from the website to identify the users who posted the content. We then applied the genderize.io API to infer the gender, following approaches used in other studies~\cite{lin2016recognizing}. However, this method has limitations outlined in previous research such as failing to identify non-binary people~\cite{noei2022study}. Other methods, such as using pronouns~\cite{Gordon2021}, were not available in our dataset.

\section{Future Work and Conclusion}

Online discussions on Workplace StackExchange offer a valuable perspective on the professional challenges and developmental needs faced by software developers. This paper provides valuable insights into the different workplace-related topics that developers discuss online. Our analysis of 47,368 posts reveals that discussions on conflicts and mistakes are the most popular, comprising approximately 30\% of the conversations. Other notable topics include workplace culture, harassment, and corporate policy issues, underscoring significant interest in these areas. We observed that corporate policy discussions tend to be more challenging. Furthermore, men contribute significantly more to these discussions than women. In future work, we plan to expand this study to workplace-related discussions on other forums. We also aim to delve deeper into the most challenging and popular topics and examine gender-based participation within specific topics.

\bibliographystyle{ieeetr} 
\bibliography{bibliography} 

\end{document}